\documentclass[aip,jap,reprint]{revtex4-2}
\pdfoutput=1
\usepackage{graphicx}
\usepackage{amssymb}
\usepackage{siunitx}
\usepackage{textcomp}

\begin{document}
\newcommand{\HeT}{$^3$He}

\title{Developing compact tuning fork thermometers for sub-mK temperatures and high magnetic fields}

\author{A.J. Woods}
\email[]{ajwoods@lanl.gov}
\affiliation{Department of Physics and National High Magnetic Field Laboratory High B/T Facility, University of Florida, Gainesville, FL 32611-8440, USA.}
\affiliation{Current address: MPA-Q, Mailstop K764, Los Alamos National Laboratory, Los Alamos, NM 87545, USA}
\author{A.M. Donald}
\affiliation{Department of Physics and National High Magnetic Field Laboratory High B/T Facility, University of Florida, Gainesville, FL 32611-8440, USA.}
\author{R. Gazizulin}
\affiliation{Department of Physics and National High Magnetic Field Laboratory High B/T Facility, University of Florida, Gainesville, FL 32611-8440, USA.}
\author{E. Collin}
\affiliation{Universit\'{e} Grenoble Alpes CNRS, Institut N\'{e}el, 25 rue des Martyrs, 28042 Grenoble Cedex 9, France.}
\author{L. Steinke}
\email[]{lucia.steinke@ufl.edu}
\affiliation{Department of Physics and National High Magnetic Field Laboratory High B/T Facility, University of Florida, Gainesville, FL 32611-8440, USA.}

\date{\today}

\begin{abstract}
There is a growing demand for experiments on calorimetric and thermal transport measurements at ultra-low temperatures below 1 mK and high magnetic fields up to 16 T. Particularly, milligram-sized solid samples are of great interest. We present the development of scalable thermometers based on quartz tuning fork resonators immersed in liquid $^3$He and adapt hydrodynamic models to provide an improved description of temperature dependence in the high viscosity regime between $1$ and $10~\si{\milli\kelvin}$. We demonstrate successful thermometer operation and discuss the feasibility of fast and compact thermal probes suitable for small samples.
\end{abstract}

\maketitle

\section{Introduction}

The NHMFL High B/T facility at the University of Florida in Gainesville pursues the mission to enable user experiments at the combined extremes of high magnetic fields (high $\mathit{B}$) and ultra-low temperatures (ULT) below 1 mK. While almost all measurement techniques like electrical transport or magnetometry require adaptations to be successfully performed in this environment, calorimetry and thermal transport are particularly challenging due to a lack of suitable thermometry. High-resolution thermal measurements in this regime would be of great benefit to the study of exotic Fermi surfaces via quantum oscillations\cite{Tan2015,Aoki2022}. Calorimetric measurements of quantum spin liquid (QSL) candidates could observe gapless excitations down to ultra-low temperatures, which would be important to demonstrate a QSL ground state\cite{Broholm2020}. Furthermore, thermal transport is a key experimental probe for the nodal structure of superconductors\cite{Matsuda2006,Sutherland2012,Metz2019}, and improved ULT thermometry could allow for studies of superconductors with extremely low transition temperatures such as bismuth \cite{Prakash2016}, as well as revealing characteristics of topologically non-trivial systems via the thermal Hall effect\cite{Kawano2019,Zhang2020}.

Motivated by these experimental challenges, we seek to develop thermometers that are suitable for measurements of typically milligram sized solid samples and are compatible with environments below 1 mK and high magnetic fields of 16 T and beyond. Existing thermometry options typically have at least one drawback that makes them unsuitable for these combined goals. Resistive thermometry has a minimum temperature of around 20 mK\cite{Courts2008}, and SQUID noise thermometers are incompatible with high magnetic fields\cite{Rothfuss2016, Kirste2016}. \HeT~ melting curve thermometry\cite{Ni1995} and nuclear magnetic resonance thermometers \cite{Tycko2013} have too high masses and long thermalization times to allow for measurements of small samples. Coulomb blockade thermometers show promise as ultra-low temperature thermometers \cite{Pekola1998,Casparis2012,Palma2017,Yurttaguel2021}, but typically require delicate on-lead magnetic demagnetization stages or on-chip cooling, which makes them incompatible or too complicated to use as sample thermometers in high magnetic fields.

Here we investigate quartz tuning fork thermometry in liquid \HeT~, which could potentially meet all of the experimental requirements of low thermal masses, sensitivity to ULT and compatibility with high magnetic fields. The measurement principle relies on the temperature dependent viscosity of liquid \HeT~ that is probed by tracking the resonance of quartz tuning forks immersed in the liquid. First, quartz tuning forks are widely used in studies of the ULT properties of the helium liquids \cite{Bradley2012,Ahlstrom2014,Jackson2015,Clubb2004,Samkharadze2011} with a minimum temperature of 100 $\mu$K in \HeT, therefore their compatibility with a ULT environment is well established. Second, thermometers tracking the vacuum resonance of a tuning fork have been used at intermediate magnetic fields up to 8 T \cite{Clovecko2019}, and magnetometers based on tuning forks in $^4$He exchange gas have even been successfully operated at high magnetic fields up to 64 T and at temperatures around 1.3 K \cite{Modic2018,Modic2020}. Third, since the tuning forks measure viscosity - an intrinsic property of the \HeT~liquid - the thermometer signal does not depend on its mass, allowing for miniaturized temperature probes for studies on small solid samples.

We report on first measurements that subject quartz tuning fork thermometers (TFTs) filled with liquid \HeT~ to the combined extremes of ultra-low temperatures down to 1 mK and high magnetic fields up to 14.5 T. We measure in the regime above the superfluid transition, where the viscosity of helium 3 is so high that simple hydrodynamic models\cite{Blaauwgeers2007,Bradley2012a} fail to predict the tuning fork resonance characteristics, and where the tuning forks require large driving forces which may affect the field dependence. In this work we attempt an adaptation of existing hydrodynamic models that take into account slip in the high viscosity regime based on previous treatments for different vibrating object geometries modified with empirical fit parameters\cite{Brumley2010, Jensen1980, Zhang2019}. We also revisit the idea\cite{Blaauwgeers2007} that even without observing the fixed point at the superfluid transition, TFTs in liquid \HeT~ may provide almost-primary thermometry with a predictable temperature dependence and require minimal or no calibration. We argue that the geometry-dependent crossover to the slip-corrected regime could serve as an alternative fixed point. Finally, we discuss recent progress in miniaturization of TFTs, and establish current resolution limits for thermal measurements based on prototype dimensions and the resolution obtained in the experiments presented here.

\section{Tuning Fork Thermometry at Ultra Low Temperature and High Magnetic Field}
We first establish the vacuum properties of the Lorentzian tuning fork resonances, then discuss the temperature dependent damping due to the viscosity of the normal fluid \HeT. We go on to compare the temperature dependence of the resonance properties to predictions of two hydrodynamic models. We also verify that the tuning forks are insensitive to magnetic fields even when immersed in highly viscous liquid \HeT, where comparatively large drive amplitudes are required.

\subsection{Tuning Fork Vacuum Properties}
\begin{figure}
\includegraphics[width=\columnwidth]{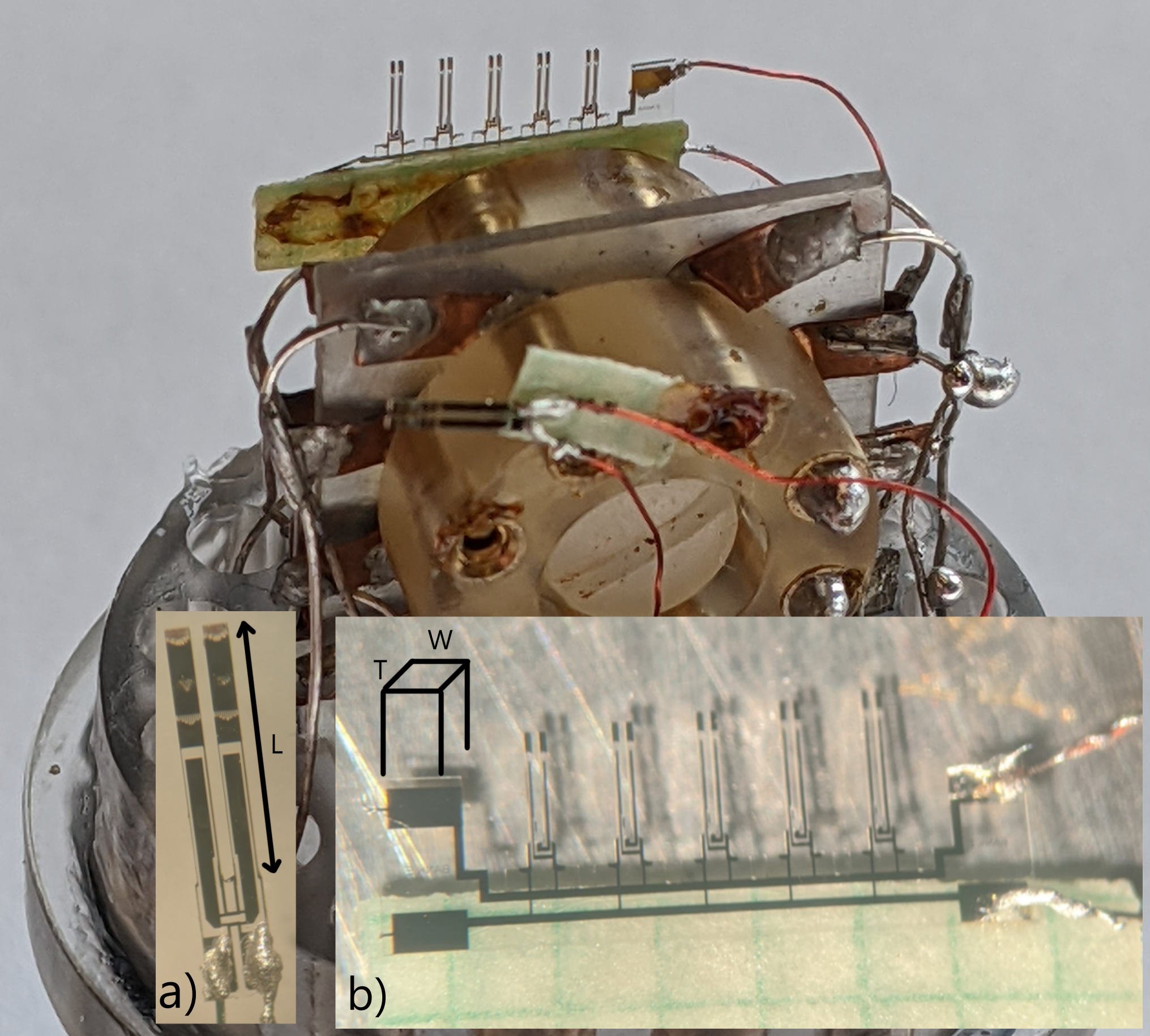}%
\caption{\label{cellForkPhoto}The polycarbonate cell insert used in this work, showing the CTF (front) and Array of tuning forks (back). Inset: the tuning forks studied in this work. a) an Epson C-002RX commercial tuning fork (CTF), nominal resonance frequency 32.768~\si{\kilo\hertz}. b) an array of 5 tuning forks, provided by Viktor Tsepelin, Lancaster University, UK\cite{Noble2022}.}
\end{figure}

Quartz tuning forks are piezoelectric resonators and are characterized by the tuning fork constant $a$, which is used to determine the velocity response to an applied driving force. $a$ depends only on the properties of the quartz and the tuning fork dimensions, the tine length, $\mathcal{L}$, the tine thickness, $\mathcal{T}$ and the tine width, $\mathcal{W}$. The tuning forks we use here are the Epson C-002RX, a commercially available device, hereafter referred to as CTF, and a tuning fork from an array manufactured by Statek corporation for use in previous work on quantum turbulence in superfluid \HeT\cite{Noble2022}. We chose the tuning fork that showed the strongest response in vacuum testing from this array, labeled ATF5. The tuning forks are shown in the inset of figure \ref{cellForkPhoto} and their physical dimensions are summarized in table \ref{dimTable}.

\begin{table}
\caption{\label{dimTable}Summary of the dimensions of the tuning forks used in this work.}
\begin{ruledtabular}
\begin{tabular}{c c c c c c}
Tuning Fork & $\mathcal{L}, \si{\micro\meter}$ & $\mathcal{T}, \si{\micro\meter}$ & $\mathcal{W}, \si{\micro\meter}$ & a, \si{\coulomb\per\meter} & $\Delta f_{vac}, \si{\hertz}$\\
CTF & 3000 & 230 & 100 & $2.78 \times 10^{-6}$ & 0.110\\
ATF5 & 1450 & 90 & 75 & $8.42 \times 10^{-7}$ & 0.070\\
\end{tabular}
\end{ruledtabular}
\end{table}

To use the tuning forks as thermometers the vacuum resonance frequency and width need to be known. We therefore took care to measure the properties in vacuum at $T \leq 1.5~\si{\kelvin}$. This temperature was chosen to minimize the residual vapor pressure in the cell, and to eliminate the temperature dependence of the quartz properties\cite{Blaauwgeers2007}.
The tuning fork constant $a$ is derived from the properties of the vacuum resonance and is given by:
\begin{equation}
a = \sqrt{\frac{4{\pi}m_{e}I_0\Delta{f}}{V_0}},
\label{ForkConstCalc}
\end{equation}
where $I_0$ is the response current peak amplitude, $\Delta{f}$ is the width of the Lorentzian resonance and $m_{e} = 0.25\rho_q\mathcal{LWT}$ is the effective mass of a tuning fork tine. Further details are given in the supplementary material. The vacuum properties of the tuning forks are summarized in table \ref{dimTable}. The measured width for the ATF5 is consistent with the ULT ($2~\si{\milli\kelvin}$) width measured previously for a similar fork\cite{Ahlstrom2014}.

\begin{figure}
\includegraphics[width=\columnwidth]{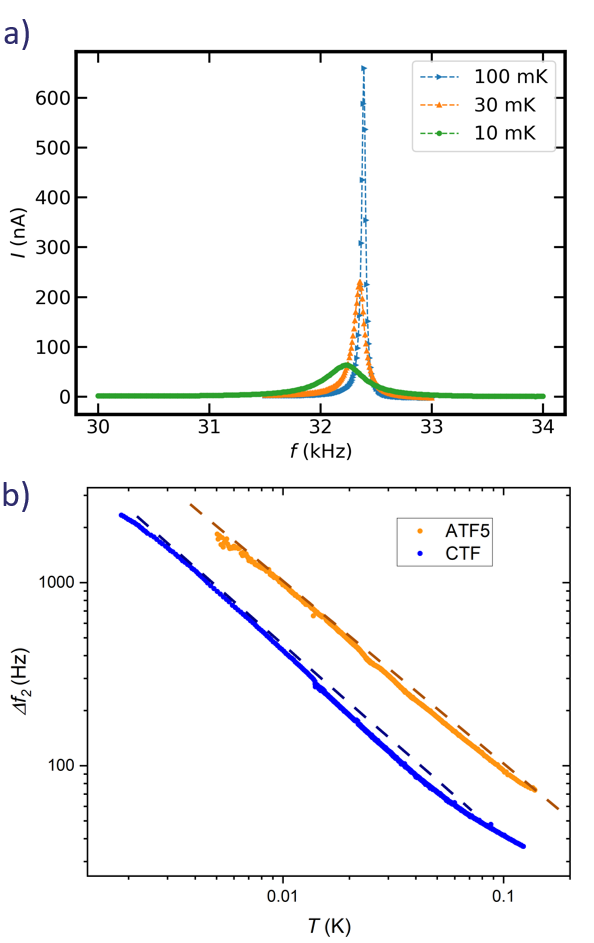}%
\caption{\label{resonanceBoth}a) Resonance curves (response current as a function of frequency) at constant driving force for the CTF b) Resonance width as a function of temperature for CTF and ATF5, the dashed lines are guides to the eye for $T^{-1}$.}
\end{figure}
\begin{table}
\caption{\label{ResTable}A summary of the resonant widths of each tuning fork for different temperatures in normal fluid \HeT.}
\begin{ruledtabular}
\begin{tabular}{c c c}
Temperature [\si{\milli\kelvin}] & CTF $\Delta{f}$ [\si{\hertz}] & ATF5 $\Delta{f}$ [\si{\hertz}]\\
100 & 40 & 90 \\
30 & 120 & 300 \\
10 & 410 & 950 \\
5 & 1065 & 1600 \\
\end{tabular}
\end{ruledtabular}
\end{table}

\subsection{Temperature Dependence}
Measurements were performed in the Bay 3 cryostat at the NHMFL High B/T Facility, at $0~\si{\bar}$ in a liquid \HeT~cell cooled by a PrNi$_5$/Cu hybrid nuclear demagnetization stage. The base temperature of this cryostat is around $700~\si{\micro\kelvin}$ and primary thermometry was provided by a \HeT~melting curve thermometer (MCT)\cite{Ni1995}. The tuning fork resonances were continuously monitored as the temperature was changed in the range  $1~\si{\milli\kelvin}$ - $100~\si{\milli\kelvin}$ by magnetizing and demagnetizing the nuclear stage.

Figure \ref{resonanceBoth}a) shows the measured resonance curves for the CTF, for three different temperatures in liquid \HeT. The widths obtained from Lorentzian fits to the resonance curves depend strongly on the temperature dependent viscosity of the $^3$He, as shown in Figure \ref{resonanceBoth}b). We calculated the expected $\mathit{T}$-dependence using an equation for the resonance width of the tuning fork $\Delta{f}$ in terms of the viscosity $\eta$ of the \HeT~derived in Blaaugweers et al.\cite{Blaauwgeers2007}:
\begin{equation}
\Delta{f} = \frac{1}{2}\sqrt{\frac{{\rho}{\eta}f_{0}}{\pi}}\mathcal{CS}\frac{(f_0/f_{vac})^2}{m_{e}},
\label{WidthVisc}
\end{equation}
where $\mathcal{C}$ is a geometric constant of order unity, which typically has a value around 0.6 for tuning forks of the type used here\cite{Bradley2012a}. $\mathcal{S} = 2\mathcal{L(T+W)}$ is the surface area of the tuning fork tines and $m_{e}$ is the effective mass of a tuning fork tine. $f_0$ and $f_{vac}$ are the resonant frequency and vacuum resonant frequency, respectively. The temperature dependence of the viscosity of~\HeT~is well established\cite{Greywall1986} as $\eta\propto T^{-2}$. We therefore expect a $T^{-1}$ dependence of the width (dashed lines in figure \ref{resonanceBoth}b). At $T>20~\si{\milli\kelvin}$ the CTF deviates from the expected temperature dependence which we attribute to poor thermalization during a relatively rapid cool down, resulting in the temperature of the MCT being lower than the temperature of the cell.

Below $5~\si{\milli\kelvin}$ for the CTF and $7~\si{\milli\kelvin}$ for ATF5, despite sufficient thermalization times, both tuning forks deviate from $T^{-1}$. This shows the limitation of the simple model developed by Blaauwgeers et al. which considers the effects of two flow fields: potential flow far away from the tuning fork tines and rotational flow near the tines, leading to an enhancement of the effective mass. This model yields a good prediction of the temperature dependence of the tuning fork width, if the viscous penetration depth $\delta$, where

\begin{equation}
\delta = \sqrt{\frac{\eta}{\rho\pi{f}}},
\label{WidthVisc}
\end{equation}

which is a measure of the extent of the flow field, is small enough that the flow fields do not interfere, which is the case as long as $\delta$ is much smaller than the relevant tuning fork dimension. At the lowest temperatures studied here, this condition is no longer fulfilled, and we observe deviation from the model prediction. We label this regime the high viscosity limit and attempt an advanced hydrodynamic description in the following section. 

\subsection{\label{slipSection} Slip Correction}
At high viscosity the liquid is no longer able to follow the motion of the vibrating object immersed in it. Instead, a non-zero velocity, or slip, occurs at the surface of the tuning fork tines. After the onset of slip, the flow fields grow more slowly with increasing viscosity and the role of interfering flow fields may be diminished. In the following we discuss a semi-empirical hydrodynamic model accounting for the effects of surface slip in tuning forks by adapting models developed to include slip corrections for vibrating objects of different geometries. This provides an improved fit to experimental data and allows for the prediction of some low temperature characteristics. We combine the hydrodynamic models derived by Zhang et al\cite{Zhang2019}, Brumley et al\cite{Brumley2010}, and Jensen et al\cite{Jensen1980}.

Zhang et al\cite{Zhang2019} describe the properties of the tuning fork in terms of hydrodynamic functions. In this model, the resonant frequency of the first vibrational mode of a quartz tuning fork is:
\begin{equation}
(2\pi{f}_0)^2 = (2\pi{f}_1)^2\left(1+\frac{\pi{\mathcal{W}}}{4\rho_{q}\mathcal{T}}\rho\Gamma_{hydro}^{r'}(2\pi{f}_1)\right),
\end{equation}
and the quality factor of that mode is:
\begin{equation}
Q_1 = \frac{2\pi{f}_1}{2\pi\Delta{f}_1} = \frac{\frac{4\rho_q{e}}{\pi\rho{l}}+\Gamma^{r'}_{hydro}}{\Gamma^{i'}_{hydro}},
\end{equation}
where $2\pi{f}_0$ is the resonant frequency in vacuum, $\mathcal{W}$ and $\mathcal{T}$ are the width and thickness of the tuning fork tines, respectively. $\rho_q$ is the density of quartz, $\rho$ is the fluid density, and $\Gamma^{r'}_{hydro}$ and $\Gamma^{i'}_{hydro}$ are the real and imaginary parts of the hydrodynamic function taken from tabulated values\cite{Brumley2010}.

Using the above expression derived in the absence of slip, we apply the slip correction developed by Jensen et al\cite{Jensen1980}, modifying the hydrodynamic function as follows:
\begin{equation}
\Gamma(2\pi{f}) - 1 = \frac{1}{\frac{1}{\Gamma_0(2\pi{f})-1}-i(\frac{\mathcal{T}/2}{\sqrt{2}\delta})(\frac{l_{slip}}{l_{slip}+\mathcal{T}/2})},
\end{equation}
where $\Gamma_0(2\pi{f})$ is the hydrodynamic function without slip correction, $l_{slip} = {const}(\frac{1+\sigma}{1-\sigma})l_{mfp}$ and $\delta$ is the viscous penetration depth. $\sigma$ is the specularity of the surface of the tuning fork and $l_{mfp}$ is the mean free path in the fluid. 

\begin{figure}
\includegraphics[width=\columnwidth]{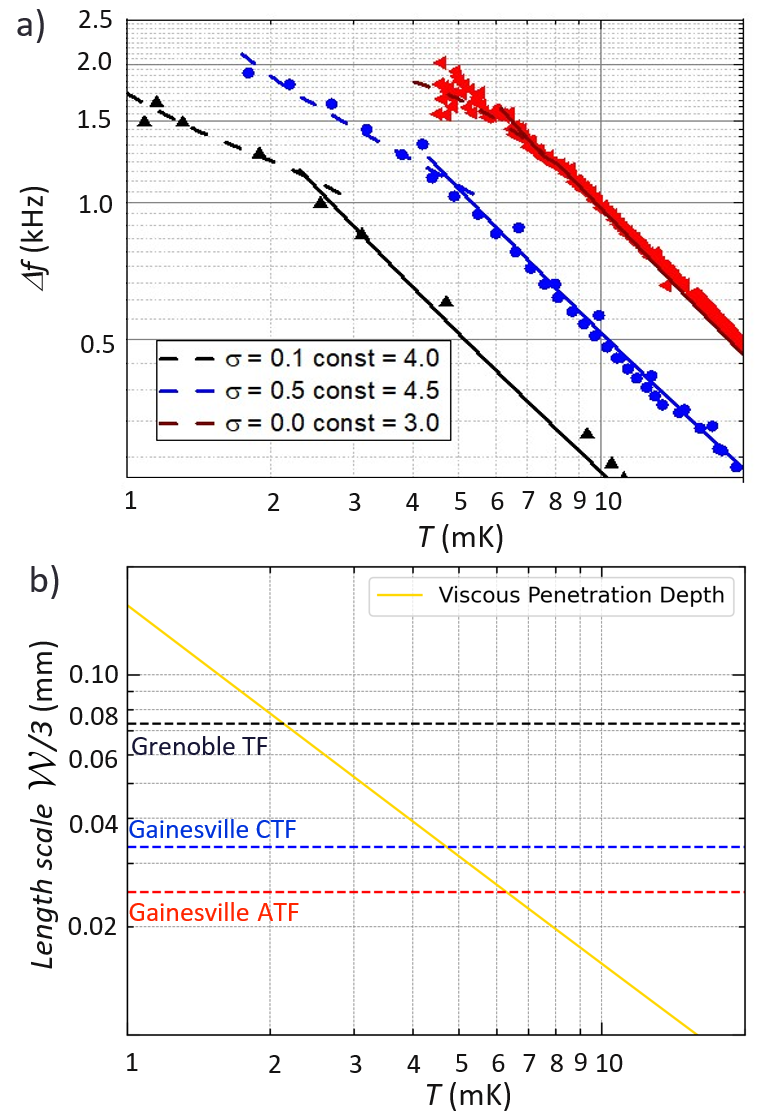}%
\caption{\label{LengthScale}a) The resonance width as a function of nuclear stage temperature for GTF(Black) CTF(Blue) and ATF5(red). The solid lines are calculated values of the width using hydrodynamic functions from Brumley et al\cite{Brumley2010}, the dashed lines are a slip correction \cite{Jensen1980} where the specularity, $\sigma$, and slip length coefficient ${const}$ are fit parameters. b) A plot showing the tuning fork length scale versus temperature. The solid line shows the viscous penetration depth, and the dotted lines the values of $\mathcal{W}/3$ for each tuning fork measured here. }
\end{figure}

We apply the modified hydrodynamic model to improve the description of the temperature dependent resonance properties of three different tuning forks. Figure \ref{LengthScale}a) shows the width of the ATF5, CTF and a third tuning fork, measured previously in Grenoble, and referred to as GTF resonances as a function of temperature in the range 20 mK to 1 mK. The dashed lines show the result of the calculation of the slip corrected tuning fork width. We allow the specularity, $\sigma$, to vary as a fork-dependent, but temperature-independent fitting parameter, which is likely related to the details of the tuning fork surface.

We observe that the slip correction becomes necessary at temperatures $T \leq T_s $ where $T_s$ is the temperature at which $\delta =\mathcal{W}/3$, as shown in figure \ref{LengthScale}b). This limit is consistent with that observed for slip in a vibrating wire resonator\cite{Winkelmann2004}. We conclude that the deviation from the simple model is indeed due to slip and suggest that this crossover could be used as a fixed point for calibrating these thermometers as the temperature where it occurs is predicted by the tuning fork geometry. 

\subsection{High Field Tests}
To establish the suitability of TFTS at high $\mathit{B}$ and ULT we performed two sets of tests: field sweeps near $18~\si{\milli\kelvin}$, the base temperature of the dilution refrigerator and demagnetization cool-downs at fixed high field. 

\begin{figure}
\includegraphics[width=\columnwidth]{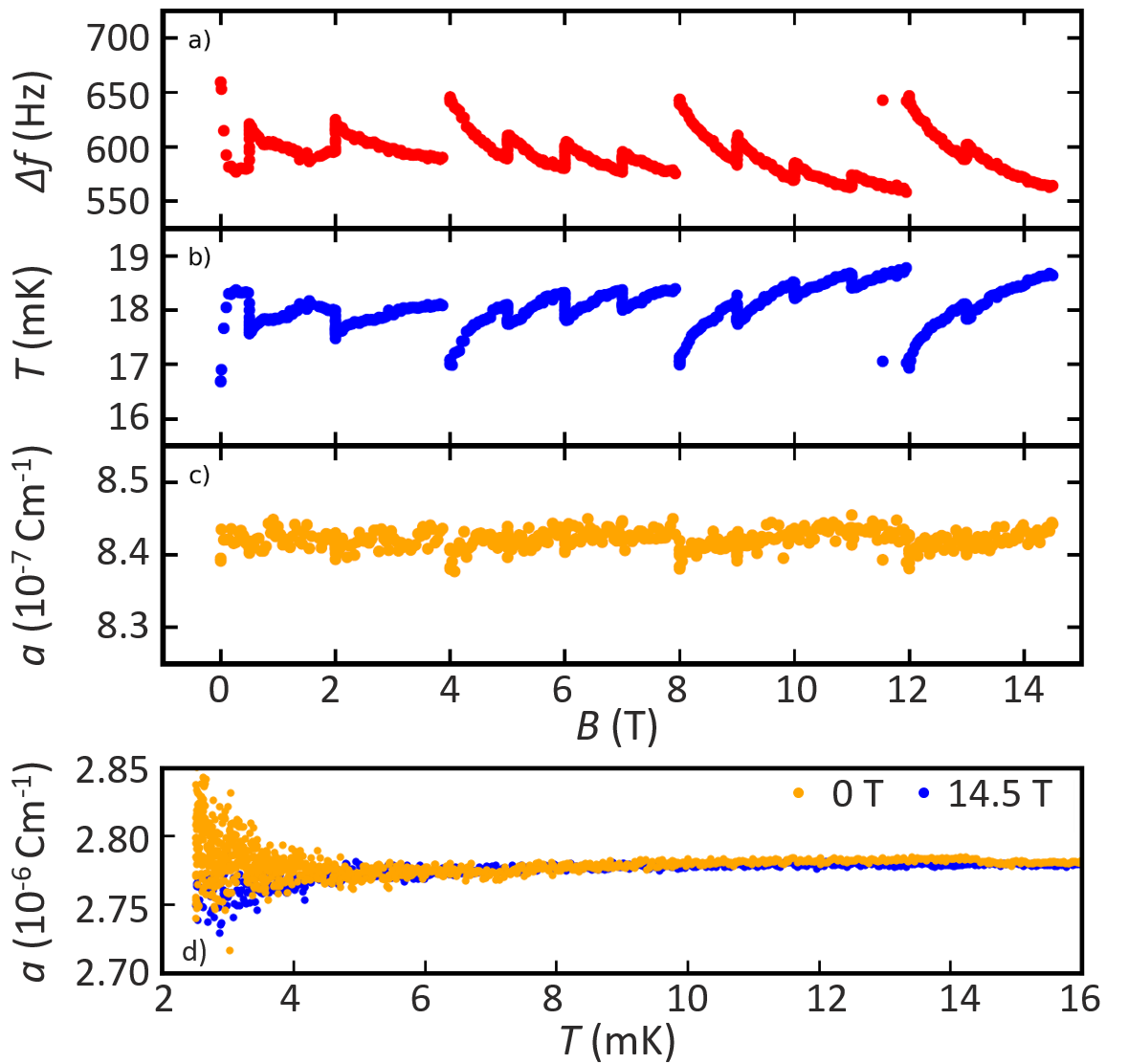}%
\caption{\label{fieldSweep} a) The resonance width of ATF5 as function of applied magnetic field during a sequence of field sweeps. b) The temperature inferred from the melting curve thermometer during the same field sweeps. c) The tuning fork constant, $a$, as calculated from the fitted resonance curve as a function of the applied field. d) The tuning fork constant, $a$ as a function of temperature for the CTF, during two different demagnetizations at $14.5~\si{\tesla}$ (blue) and $0~\si{\tesla}$ (orange)}
\end{figure}

Figure \ref{fieldSweep}a) shows the width of ATF5 as a function of the applied magnetic field. The width changes as the field is changed, and relaxes while the field remains constant. We attribute this to eddy current heating in the silver cold finger that links the nuclear stage to the \HeT~cell, causing heating of the \HeT~in the cell, consistent with the temperature measured indpendently by the MCT (\ref{fieldSweep}b)). Figure \ref{fieldSweep}c) shows the tuning fork constant, $a$, as a function of the applied field. While the tuning fork resonance shows a direct response to the temperature changes in the cell, the tuning fork constant remains within 0.6 \% of its $B=0$ value even at the highest field of $14.5~\si{\tesla}$, confirming that there is no residual magnetic field dependence once the temperature changes are accounted for. This illustrates that ULT cannot maintained during a magnetic field sweep, therefore performance tests at ULT and high $\mathit{B}$ are achieved by $\mathit{T}$-sweeping at constant magnetic field.

Figure \ref{fieldSweep}d) shows the tuning fork constant, $a$, of the CTF as function of temperature at $14.5~\si{\tesla}$ (blue) and $0~\si{\tesla}$ (orange) where the zero-field fork constant is derived from the data presented in figure \ref{resonanceBoth}b). The CTF exhibits the same insensitivity to the applied magnetic field as the ATF5. This confirms that there is no significant magnetic field induced effect on the properties of the quartz, the tuning fork electrodes or on the viscosity of the $^3$He. The tuning forks are therefore suitable as thermometers in the temperature and magnetic field ranges of interest. The deviation of $a$ below $5~\si{\milli\kelvin}$ results from the onset of slip, and further work is required to correct the tuning fork constant in this regime.

\section{Tuning forks in \HeT~as thermal probes for small samples}

\begin{figure}
\includegraphics[width=\columnwidth]{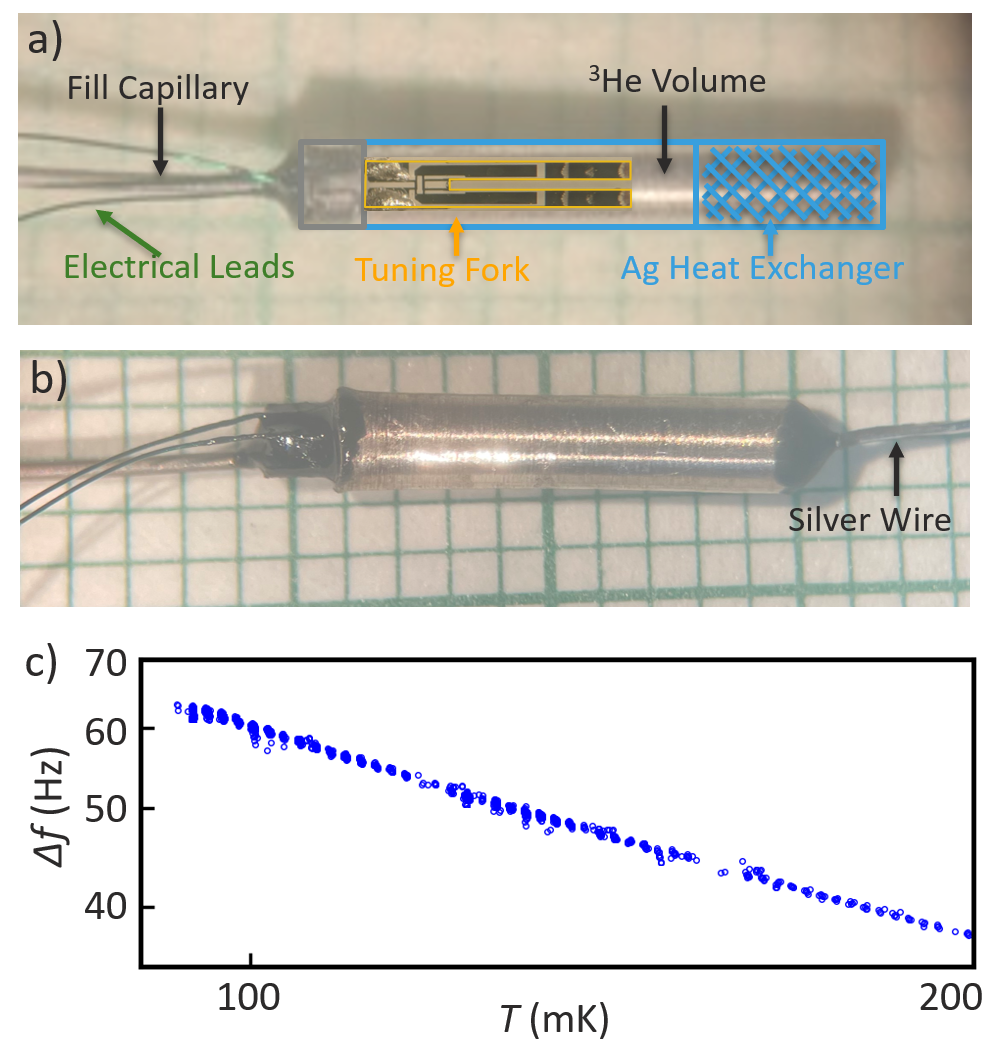}%
\caption{\label{TFTPhoto}Prototype tuning fork thermometers comprising a CTF and silver sinter heat exchanger encased in a silver tube. a) Schematic overlay showing the internal components and dimensions b) a newer prototype, showing the silver wire for the thermal link to the sample c) The resonance width as a function of temperature for the TFT pictured in b).}
\end{figure}

Having established compatibility of the TFTS with high $\mathit{B}$ and ULT, we discuss adaptations for the measurement of small solid samples. To probe such samples in specific heat and thermal transport measurements, the thermal mass of the thermometer must be small enough that temperature changes caused by the thermal response of the sample can be detected. A lower thermal mass also corresponds to a faster response time of the thermometer, which allows for measurements with higher resolution in a given time frame. This can be achieved by reducing the \HeT~volume as much as possible, since the dominant contribution to the ULT specific heat of \HeT~cells generally stems from the liquid itself\cite{Greywall1986}. As the tuning fork directly probes the viscosity of the surrounding liquid, its measurement signal is independent of the total \HeT~mass. Therefore, reducing the amount of \HeT~should not come at the expense of temperature resolution or alter the measurement result. However, if the tuning fork comes within the viscous penetration depth of the wall of the thermometer, additional adaptions to the hydrodynamic model will be required to correctly predict the temperature dependence. As the thermal mass of the thermometer is reduced, care should also be taken to maintain the tuning fork driving force and tine velocity at a low enough level that the tuning fork does not self heat the thermometer, or to run the tuning fork in pulses so as to reduce the time that the tuning fork can deposit heat into the thermometer.

Figure~\ref{TFTPhoto} shows prototype tuning fork thermometers that represent first attempts at minimizing the \HeT~volume while still allowing for enough space for the tuning for to ring freely, to add or remove \HeT~via a fill capillary, and ensuring good thermal contact via an integrated heat exchanger. Tuning forks have previously been used in compact geometries for bolometry in superfluid \HeT-B\cite{Noble2022} and for thermal transport in confined \HeT\cite{Lotnyk2020}. The key difference here is the complete enclosure with a heat exchanger and with the intention of mounting a small solid state sample for thermal measurements. Both prototypes contain nominally identical CTFs and silver heat exchangers and are sealed with epoxy plugs at either end. The thermometer in panel a) is constructed from a high purity silver body and a schematic of the components inside is overlayed, b) shows a variation on this design, with a silver wire directly pressed into the heat exchanger, to improve thermal contact with a sample mounted on the wire. We estimate the total specific heat of these thermometers to be around $4~\si{\micro\joule\per\kelvin}$ at $1~\si{\milli\kelvin}$, based on published values for \HeT\cite{Greywall1986}~and silver\cite{Martin1973}. The thermalization time of the thermometer depends on the specific heat and the thermal resistance determined by the surface area of the silver heat exchanger\cite{Pollanen2009}. For a heat exchanger of the size shown in figure \ref{TFTPhoto}a), assuming a 50 \% packing fraction, we calculate that the response time of the thermometer is of order 50 s, which is a comparably rapid response in the context of ULT experiments. Figure \ref{TFTPhoto}c) shows the resonance width of the TFT pictured in figure \ref{TFTPhoto}b) down to temperatures of around $90~\si{\milli\kelvin}$, showing comparable temperature dependence to tuning forks in larger cells, which confirms adequate thermalization and establishes that mini TFTs of this design are leak tight and can be successfully filled by condensing \HeT. The next step is to test this thermometer and other prototypes down to $1~\si{\milli\kelvin}$.

Based on the present accuracy of the temperature measurement based on the tuning fork width ($\approx 50 \si{\micro\kelvin}$), we assume that the detection limit of the tuning fork thermometer is around 1\% of its own specific heat, i. e. it should be capable of measuring samples with specific heat $40~\si{\nano\joule\per\kelvin}$, comparable to the specific heat of small single crystal samples that users of the High B/T facility may want to investigate. 
\section{Summary}
We have shown that quartz tuning forks immersed in liquid \HeT~provide robust thermometry at the combined extremes of ultra-low temperatures down to 1 mK and high magnetic fields up to 14.5 T, presenting a viable platform for temperature measurements in this regime. We also presented an adaption of hydrodynamic models to improve temperature calibration in the high viscosity regime between $1$ and $10~\si{\milli\kelvin}$. We discussed the feasibility of miniature thermometers that could provide much needed thermal probes for user experiments on small solid state samples in the extreme environments accessible at the NHMFL High B/T Facility. Existing thermometer prototypes built with relatively simple fabrication methods could already be suitable for a large number of measurements extending the typical temperature range for thermal measurements below $\sim$ 20 mK.

\begin{acknowledgments}
Experiments were performed at the National High Magnetic Field Laboratory (NHMFL) High B/T Facility at the University of Florida and at Universit\'{e} Grenoble Alpes CNRS.
The NHMFL is supported by the National Science Foundation Cooperative Agreement DMR-1644779 and the State of Florida.
The authors also acknowledge support from the NHMFL User Collaboration Grants Program (UCGP).
The Grenoble ULT Group has received funding from the European Union's Horizon 2020 Research and Innovation Programme under grant agreement No. 824109, the European Microkelvin Platform (EMP).
The quartz tuning fork array was provided by Viktor Tsepelin, Lancaster University, UK.
AJW, AMD, RG, EC and LS took the data, AJW, AMD, RG, EC and LS analyzed the data and AJW and LS wrote the manuscript.
\end{acknowledgments}

\bibliography{ForkPaperBib}

\end{document}